\documentstyle[12pt]{article}

\begin{document}

\setcounter{page}{100}

\centerline{\large {\bf Wave Function Properties in a High
Energy Process}}
\vskip.3in
\centerline{ Arjun Berera}
\centerline{ Department of Physics}
\centerline{ Pennsylvania State University}
\centerline{ University Park, Pennsylvania 16802, USA}

\vskip.6 in
\centerline{\bf Abstract}
\vskip.1in
A model example is given of how properties of the
hadronic light-cone wave function are revealed
in a particular high energy process.
\vskip.1in

\vskip.3in
{\bf 1. Introduction}
\vskip.1in

The purpose of my talk is to give a pedagogical derivation
of the light-cone wave function and demonstrate its use in a
particular high energy process.
Explicitly I will derive the meson wave function
in scalar quark QCD.  I will then apply it to recent work I have been
doing with David Soper on diffractive hard scattering.

The Lagrangian for scalar quark QCD coupled to a meson field is,
\begin{eqnarray}
{\cal L} &=&  (D_{\mu}q)^\dagger ( D^{\mu}q) -m^2 q^\dagger q
-{1\over 4} G^{a}_{\mu \nu }G_{a}^{\mu \nu }
-{1\over 2\xi}(\partial_\mu A_a^\mu)(\partial_\nu A_a^\nu)
\nonumber\\
&&+ \ \hbox{\rm Faddeev-Popov terms}
-{g_4\over 4}(q^\dagger q)^2.
\end{eqnarray}
Here $A^\mu_a(x)$ as an SU(3) gauge field
as in normal QCD. There is also a color triplet quark
field $q_i(x)$ with quark mass m, but we take $q_i(x)$ to be a scalar
field instead of a Dirac field. Since the theory includes a scalar
quark field, a 4-quark coupling is necessary, but we have set the
renormalized coupling constant $g_4$ to a negligible small value.
This theory has the same behavior as spinor QCD for collinear and
soft gluon emission from quarks.  Its chief advantage is that it
allows a perturbative model for a quark-antiquark bound state.  We
introduce a scalar, color singlet meson field $\phi(x)$ and couple it
to the quarks using
\begin{equation}
{\cal L}_\phi = G\ \phi(x)\, q^\dagger_i(x) q_i(x).
\end{equation}
We work to lowest nontrivial order in the $\phi q^\dagger q$ coupling
$G$, letting the $\phi q^\dagger q$ vertex play the role that is played
by the (amputated) Bethe-Salpeter wave function of a meson in spinor
QCD. We denote the mass of the meson by $M$ and take $M$ to be smaller
than $2m$, so that the meson cannot decay into a quark and an
antiquark.

\vskip.3in
{\bf 2. Light-Cone Wave Function}
\vskip.1in

For a meson moving in the
minus direction, the light-cone wave function $\psi(x,{\bf k})_{ij}$ is the
amplitude to find that the meson with momentum $P^\mu = (M^2/(2P^-),
P^-,{\bf 0})$ consists of a quark and an antiquark of colors $i$ and
anti-$j$ respectively, with the quark having minus-momentum $k^- =
xP^-$ and transverse momentum ${\bf k}$.  The wave function is
measured by operators defined on the null-plane $y^- = 0$.  The
precise definition, following the formalism of [1,2,3]
is,
\begin{equation}
\psi(x,{\bf k})_{ij} =
2x(1-x)P^- \int d^4 y\ e^{ik\cdot y}\, \delta(y^-)\
\langle 0|
q_i(y)\,q^\dagger_j(0)
|P\rangle \,.
\label{psidef}
\end{equation}
Here we have chosen the normalization
\begin{equation}
(2\pi)^{-3}\int_0^1 {dx \over 2x(1-x)}\int d{\bf k} \sum_{ij}
|\psi(x,{\bf k})_{ij}|^2
= P_{\!2}\,,
\label{psinorm}
\end{equation}
where $P_{\!2}$ is the probability, which is of order $G^2$, that the
meson state consists of a $(q,q^\dagger)$ pair.  In terms of the
covariant $\phi\, q\, q^\dagger$ Green function amputated on the
$\phi$-leg, the definition (\ref{psidef}) can be written as
\begin{equation}
\psi(x,{\bf k})_{ij} =
2x(1-x)P^- \int {\ dk^+\over 2 \pi}\, {\cal
G}(k^\alpha,P^\beta)_{ij}\,.
\label{psigreen}
\end{equation}
At lowest order in $\alpha_s$ and $G$ one has,
\begin{equation}
{\cal G}(k^\alpha,P^\beta)_{ij} = iG\,\delta_{ij}\,
{i \over k^2 - m^2 + i\epsilon}\
{i \over (P-k)^2 - m^2 + i\epsilon}
\,.
\end{equation}
By integrating according to Eq.~(\ref{psigreen}), we find
\begin{equation}
\psi(x,{\bf k})_{ij} =
{ G x(1-x)
 \over {\bf k}^2 + m^2 - x(1-x)M^2}
\ \delta_{ij}\,.
\label{psi}
\end{equation}
Notice that $|\psi|^2 \propto 1/{\bf k}^4$ for large ${\bf k}^2$.
This good behavior in the ultraviolet, which arises from the fact that
$G$ has dimensions of mass, is the essential reason for the
usefulness of this model.

The wave function can be used to calculate, to order zero in
$\alpha_s$, the probability for finding a quark in a meson
{\it i.e.} the parton distribution function:
\begin{eqnarray}
f_{q/\phi}(x) &=&
(2\pi)^{-3}{1 \over 2x(1-x)}\int d{\bf k} \sum_{ij}
|\psi(x,{\bf k})_{ij}|^2
\nonumber\\ &=&
{3 G^2\over 16 \pi^2}
{ x(1-x)
\over m^2 - x(1-x)M^2 }\,.
\label{pdf}
\end{eqnarray}

As a more realistic example, one can derive the light-cone wavefunction
for quarks in a photon.
For transverse
polarization, we find
\begin{equation}
\psi(x,{\bf k}) =
-{e{\cal Q} \over {\textstyle 2P^-}}\
{\overline U\ \left[
x\ \epsilon\cdot\gamma\ k_T\cdot\gamma
- (1-x)\ k_T\cdot\gamma\ \epsilon\cdot\gamma
\right]\gamma^- V
\over {\bf k}^2 + x(1-x)Q^2},
\label{psiT}
\end{equation}
where $k_T^\mu = (0,0,{\bf k})$. This is quite similar to the scalar
wave function, Eq.~(\ref{psi}).  The chief difference is the factor
of ${\bf k}$ in the numerator, which leads to a logarithmic
divergence in the normalization integral for $\psi$. (The spinors $U$
and $V$ depend on ${\bf k}$, but this dependence is eliminated when
the spinors stand next to a $\gamma^-$.) For longitudinal
polarization, we obtain
\begin{equation}
\psi(x,{\bf k}) =
{e{\cal Q} \over {\textstyle P^-}}\
x(1-x)Q\
{\overline U\gamma^- V
\over {\bf k}^2 + x(1-x)Q^2}\,.
\label{psiL}
\end{equation}
Again, $\overline U\gamma^- V$ is independent of ${\bf k}$. This
wave function is small compared to that for transverse polarization
when $Q^2 \ll {\bf k}^2$ but becomes comparable when $Q^2 \sim {\bf
k}^2$.

\vskip.3in
{\bf 3. Diffractive Hard Scattering}
\vskip.1in

In 1985 Ingelmann and
Schlein$^4$
predicted that events of the type
\begin{equation}
A+B \to A + {\rm jets} + X,
\label{dhs}
\end{equation}
where hadron $A$ is diffractively scattered, should occur with a small
but not tiny probability.  Here by ``diffractively scattered,'' we mean
that $A$ emerges with a fraction $(1 - z) > 0.9$ of its original
longitudinal momentum and with a small transverse momentum
$|{\bf P}_{\!A}^{\prime}| \leq 1 {\rm\ GeV}$.  The transverse
momentum transfer can also be characterized using the invariant
momentum transfer $t$ from the hadron: $t = (P_{\!A} -
P_{\!A}^\prime)^2 = - ({\bf P}_{\! A}^{\prime\, 2} + z^2 M_A^2) /
(1-z) \approx - {\bf P}_{\! A}^{\prime\, 2}$.

The picture for such diffractive hard scattering proposed by Ingelman
and Schlein is that hadron $A$ exchanges a pomeron with the rest of
the system, where ``pomeron'' means whatever is exchanged in elastic
scattering at large $s$, small $t$. Thus the cross section is
proportional to the pomeron coupling to hadron $A$ as measured in
elastic scattering. The pomeron carries transverse momentum $-{\bf
P}_{\! A}^{\prime}$ and a fraction $z$ of the hadron's longitudinal
momentum.  Here we do not need to know what a pomeron is, only that
its momentum is carried by quarks and gluons. One of these collides
with a parton from hadron $B$ to produce the jets.  Let the parton
that participates in the hard scattering carry a fraction $x$ of the
longitudinal momentum of the incoming hadron $A$, and thus a fraction
$x/z$ of the longitudinal momentum transferred by the pomeron.  Then
the cross section in this model is proportional to a function
$f_{a/P}(x/z,t;\mu)$, where $f_{a/P}(\xi,t;\mu)\, d\xi$ is interpreted
as the probability to find a parton of kind $a$ in a pomeron, where
the parton carries a fraction $\xi$ of the pomeron's longitudinal
momentum.

The reaction (\ref{dhs}) anticipated by Ingelman and Schlein has been
seen at the CERN collider by the UA8 experiment$^5$.
However,
the experiment suggests a feature not anticipated in [4,6,7].
It was expected that the functions
$f_{a/P}(x/z)$ would have support only for $x<z$.  That is, some of
the momentum fraction $x$ transferred from hadron $A$ would be lost,
appearing in low $P_T$ particles rather than in the jets.  Instead, the
experiment suggests that a fraction of the events are lossless in the
sense that $x=z$.  It is as if the formula for the cross section
contained a term proportional to $\delta(1-x/z)$.  A similar such
distributional form was predicted in [8].

We will consider the cross section for lossless jet production
in diffractive hard scattering.
The details of the calculation are given in [9].
Our purpose here is to examine the role played by the light-cone
meson wavefunction.
The cross
section is,
\begin{eqnarray}
\left[{d \sigma^{\rm diff}(A + B\to A +{\rm jets} +X)\over
d E_T\, dX_A\, dX_B\, dz\, dt}\right]_{0}
&\sim&
\delta\left( 1-X_A/z  \right)
\int d{\bf r}\,
{|\psi(X_B,{\bf r})|^2 \over 2X_B(1-X_B)}
\nonumber\\ &&\hskip -4cm \times
\sum_{j,k = 1}^2 \sum_{a,b = 1}^8 {\rm Tr} \biggl\{
\left[ G^j_a(-{\bf r};t,z) - G^j_a({\bf 0};t,z)
\right]^\dagger H^{jk}_{ab}\left(\hat s,E_T\right)
\nonumber\\ && \times
\left[ G^k_b(-{\bf r};t,z) - G^k_b({\bf 0};t,z)
\right]\biggr\}.
\label{result}
\end{eqnarray}
%

%

Despite its rather complicated structure, the interpretation of
Eq.~(\ref{result}) is straightforward. In the model, meson $B$
consists of a quark and an antiquark.  With probability $\propto
|\psi(X_B,{\bf r})|^2$, they are separated by a transverse distance
$\bf r$. In order to restore the color of hadron $A$, we must absorb a
gluon on either the antiquark (at position $-{\bf r}$) or the quark
(at position $\bf 0$). Since the quark and antiquark have opposite
color charges, the absorption amplitude is proportional to the
difference $G^j_a(-{\bf r};t,z) - G^j_a({\bf 0};t,z)$. Here
$G^j_a({\bf b};t,z)$ is the amplitude to absorb a color field
quantum at transverse position ${\bf b}$ when the ``active'' gluon is
annihilated at the origin of space-time and hadron $A$ is
diffractively scattered.  Thus $G$ describes the color field
associated with the pomeron when one gluon from the pomeron has been
annihilated at the origin.

Here we meet an interesting experimental possibility.  The ${\bf b}$
dependence of $G^j_a({\bf b};t,z)$ reflects the transverse structure
of the pomeron. It has significant structure on some distance scale
$R_{\rm P}$ characteristic of the pomeron. In the present
model, $1/R_{\rm P}$ is of order of the quark mass $m$. Thus
$G^j_a(-{\bf r};t,z) - G^j_a({\bf 0};t,z)$ is small when $|{\bf r}|
\ll R_{\rm P}$.  On the other hand, $|\psi(X_B,{\bf r})|^2$ is small
when $|{\bf r}| \gg R_B$, where $R_B$ is a characteristic size of
hadron $B$.  This size is also of order $1/m$ in the model. However,
suppose that we generalize the model so that $R_B$ can be separately
adjusted.  Then when $R_B \sim  R_{\rm P}$, there will be a
substantial contribution to the cross section proportional to
$\delta\left( 1-X_A/z \right)$.  But when $R_B \ll  R_{\rm P}$, this
contribution will vanish.

So far, we have worked only with a simple model.  But the model
suggests a plausible conjecture.  First, there can be a sizable
contribution to diffractive jet production proportional to
$\delta\left( 1-X_A/z \right)$, arising from using one gluon from the
pomeron to make the jets and absorbing  on the partons of hadron $B$
the rest of the color field needed to make hadron $A$ back into a
color singlet. Second, when the size $R_B$ of hadron $B$ is small
compared to the transverse size $R_{\rm P}$ associated with the color
field in pomeron exchange, then hadron $B$ should act as a color
singlet and this contribution should disappear.

In order to test this conjecture, and probe the transverse structure of
the pomeron, one needs to use hadrons of adjustable size.
At HERA, one manufactures bremsstrahlung
photons from the electron beam.  The virtuality  $Q = [{-P_{\!B}^\mu
P_{\!B\mu}}]^{1/2}$ of the photon is measured by the deflection of the
electron, and can be anything from nearly zero to many GeV. The photon
can collide with a proton (hadron $A$) to make jets with
$E_T\gg Q$. The cross section for this process can be (roughly) divided
into two parts. In one part, the photon acts as a parton and scatters
directly with a parton from hadron $A$ to make the jets.  In the other
part, the photon acts as a hadron, made of constituent partons. For $Q
\approx 0$, this hadron is essentially a $\rho$-meson, with a size $R_B
\approx 1 {\ \rm fm}$.  For $Q \gg 1 {\ \rm fm}^{-1}$, the ``hadron''
consists of a quark-antiquark pair, with wave functions given in
Eq.~(\ref{psiT}) for transverse polarization and Eq.~(\ref{psiL}) for
longitudinal polarization. These wave functions are characterized by a
size $R_B \approx [X_B (1-X_B) Q^2]^{-1/2}$. Since $Q^2$ and
$X_B$ are measurable, this size is adjustable.



We must emphasize that the proposal given above is a conjecture based
on a simple model, not a proven consequence of QCD.  It should be a
challenge to investigate the structure of diffractive hard scattering
further and to discover what features of the model survive a higher
order analysis.

\vskip.3in
{\bf Acknowledgment}
\vskip.1in

I thank R. Perry and S. Glazek for inviting me to this meeting.
Partial funding was provided by the U. S. Department of Energy
grant DE-FG03-91ER40674.

\vskip.3in
{\bf References}
\vskip.1in

1. J.\ B.\ Kogut and D.\ E.\ Soper, Phys.\ Rev.\
{\bf D1}, 2901 (1970).

2. J.\ D.\ Bjorken, J.\ B.\ Kogut and D.\ E.\ Soper,
Phys.\ Rev. {\bf D3}, 1382 (1971).

3. S.\ J.\ Brodsky and G.\ P.\ Lepage, Phys.\
Rev. {\bf D22}, 2157 (1980).

4. G.\ Ingelman and P. Schlein, Phys.\ Lett. {\bf B152},
256 (1985).

5.  A.\ Brandt, et. al. , Phys.\ Lett.\ {\bf B297}, 417
(1992).

6.  H.\ Fritzsch and K.\ H.\ Streng,
Phys.\ Lett.\ {\bf B164}, 391 (1985).

7.  E.\ L.\ Berger, J.\ C.\ Collins, D.\ E.\ Soper, and
G.\ Sterman, Nucl.\ Phys.\ {\bf B286}, 704

(1987).

8. L. Frankfurt and M. Strikman, Phys.\ Rev.\ Lett. {\bf 64}, 1914 (1989).

9.  A. Berera and D. E. Soper, Phys. \ Rev.  {\bf D50},
in press 1994.

\end{document}